\documentstyle[floats,epsf,twocolumn,prd,aps]{revtex}

\begin{document}

\draft
%
%
\newcommand{\nc}{\newcommand}
\nc{\bea}{\begin{eqnarray}}
\nc{\eea}{\end{eqnarray}}
\nc{\beq}{\begin{equation}}
\nc{\eeq}{\end{equation}}

\nc{\half}{\frac{1}{2}}
\nc{\EH}{{}^3{\rm H}}
\nc{\EHe}{{}^3{\rm He}}
\nc{\UHe}{{}^4{\rm He}}
\nc{\GLi}{{}^6{\rm Li}}
\nc{\ZLi}{{}^7{\rm Li}}
\nc{\ZBe}{{}^7{\rm Be}}
\nc{\DH}{{\rm D}/{\rm H}}
\nc{\EHeH}{^3{\rm He}/{\rm H}}
\nc{\GLiH}{{}^6{\rm Li}/{\rm H}}
\nc{\ZLiH}{{}^7{\rm Li}/{\rm H}}

\nc{\etal}{{\it et al. }}
\nc{\ve}[1]{{\bf #1}}
\nc{\sigt}{\sigma^{\rm t}}

%
%

\nc{\AJ}[3]{{Astron.~J.\ }{{\bf #1}{, #2}{ (#3)}}}
\nc{\anap}[3]{{Astron.\ Astrophys.\ }{{\bf #1}{, #2}{ (#3)}}}
\nc{\ApJ}[3]{{Astrophys.~J.\ }{{\bf #1}{, #2}{ (#3)}}}
\nc{\apjs}[3]{{Astrophys.~J.\ Supp.\ S.\ }{{\bf #1}{, #2}{ (#3)}}}
\nc{\apjl}[3]{{Astrophys.~J.\ Lett.\ }{{\bf #1}{, #2}{ (#3)}}}
\nc{\app}[3]{{Astropart.\ Phys.\ }{{\bf #1}{, #2}{ (#3)}}}
\nc{\araa}[3]{{Ann.\ Rev.\ Astron.\ Astrophys.\ }{{\bf #1}{, #2}{ (#3)}}}
\nc{\arns}[3]{{Ann.\ Rev.\ Nucl.\ Sci.\ }{{\bf #1}{, #2}{ (#3)}}}
\nc{\arnps}[3]{{Ann.\ Rev.\ Nucl.\ and Part.\ Sci.\ }{{\bf #1}{, #2}{ (#3)}}}
\nc{\MNRAS}[3]{{Mon.\ Not.\ R.\ Astron.\ Soc.\ }{{\bf #1}{, #2}{ (#3)}}}
\nc{\mpl}[3]{{Mod.\ Phys.\ Lett.\ }{{\bf #1}{, #2}{ (#3)}}}
\nc{\Nat}[3]{{Nature }{{\bf #1}{, #2}{ (#3)}}}
\nc{\ncim}[3]{{Nuov.\ Cim.\ }{{\bf #1}{, #2}{ (#3)}}}
\nc{\nast}[3]{{New Astronomy }{{\bf #1}{, #2}{ (#3)}}}
\nc{\np}[3]{{Nucl.\ Phys.\ }{{\bf #1}{, #2}{ (#3)}}}
\nc{\pr}[3]{{Phys.\ Rev.\ }{{\bf #1}{, #2}{ (#3)}}}
\nc{\PRC}[3]{{Phys.\ Rev.\ C\ }{{\bf #1}{, #2}{ (#3)}}}
\nc{\PRD}[3]{{Phys.\ Rev.\ D\ }{{\bf #1}{, #2}{ (#3)}}}
\nc{\PRL}[3]{{Phys.\ Rev.\ Lett.\ }{{\bf #1}{, #2}{ (#3)}}}
\nc{\PL}[3]{{Phys.\ Lett.\ }{{\bf #1}{, #2}{ (#3)}}}
\nc{\prep}[3]{{Phys.\ Rep.\ }{{\bf #1}{, #2}{ (#3)}}}
\nc{\RMP}[3]{{Rev.\ Mod.\ Phys.\ }{{\bf #1}{, #2}{ (#3)}}}
\nc{\rpp}[3]{{Rep.\ Prog.\ Phys.\ }{{\bf #1}{, #2}{ (#3)}}}
\nc{\ibid}[3]{{\it ibid.\ }{{\bf #1}{, #2}{ (#3)}}}

\wideabs{
\title{Inhomogeneous Big Bang Nucleosynthesis and Mutual Ion Diffusion}

\author{Elina Keih\"anen \cite{maile}}
\address{Department of Physical Sciences, University of Helsinki,
         P.O.Box 64, FIN-00014 University of Helsinki, Finland}


\maketitle

\begin{abstract}

We present a study of inhomogeneous big bang nucleosynthesis with emphasis 
on transport phenomena. We combine a hydrodynamic treatment to a nuclear reaction network and compute
the light element abundances for a range of inhomogeneity parameters. We find that shortly after
annihilation of electron-positron pairs, Thomson scattering on background photons prevents the
diffusion of the remaining electrons.  Protons and multiply charged ions then tend to diffuse
into opposite directions so that no net charge is carried. Ions with
$Z>1$ get enriched in the overdense regions, while protons diffuse out into regions of lower
density. This leads to a second burst of nucleosynthesis in the overdense regions at $T<20$ keV,
leading to enhanched destruction of deuterium and lithium.
We find a region in the parameter space at $2.1\times10^{-10}<\eta<5.2\times10^{-10}$ where
constraints $\ZLiH<10^{-9.7}$ and $\DH<10^{-4.4}$ are satisfied simultaneously.

\end{abstract}

}

%
%


\section{Introduction}

Inhomogeneous big bang nucleosynthesis (IBBN) has been studied in several papers
\cite{Applegate85,AHS87,Alcock87,Malaney88,Terasawa89,HX88,HX89,HX90a,HX90b,Mathews90,Mathews93,Jedamzik94b,Mathews96,Orito97,HX97,Kainulainen99,HX01,Jedamzik01}.
In IBBN the baryon density is assumed to be inhomogeneous during nucleosynthesis.
The inhomogeneity could be the result of a first-order phase transition
occurring before BBN, or of some unknown physics possibly connected
with inflation.

The effects on light element production depend strongly on the length 
scale of the inhomogeneity. It is well known that there is a so-called ``optimal scale'', at which
the production of $\UHe$ is reduced with respect to the homogeneous case, due to differential
diffusion of protons and neutrons.

The first studies on IBBN concentrated on the reduced $\UHe$ production and disregarded dissipative
phenomena other than diffusion.
Later works consider also other transport phenomena. The collective
hydrodynamic expansion of the high-density regions was first addressed by Alcock \etal
\cite{Alcock90}. Jedamzik and Fuller \cite{Jedamzik94a} give a detailed study of
dissipative processes at temperatures ranging from $T\approx100$ GeV to $T\approx 1$ keV, including
diffusion, hydrodynamic expansion, and photon inflation.

The mutual diffusion of isotopes, however, has to our
knowledge not been properly accounted for previously.  Diffusion of one ion species is not
restricted by collisions with another species, if both are moving into same direction with 
same fluid velocity. On the other hand, momentum transfer is enhanched between two fluid components
flowing into opposite directions.

Accurate treatment of transport phenomena has become increasingly important, since several 
estimations on the primordial $\ZLi$ abundance indicate a low primordial $\ZLiH$, which is difficult
to accomodate in standard big bang nucleosynthesis (SBBN). Lithium is produced quite late in
nucleosynthesis, and its yield is therefore particularly sensitive to the late-time transport
phenomena such as ion diffusion and hydrodynamic expansion of the overdense regions.

In this work we study inhomogeneous big bang nucleosynthesis  with emphasis on ion transport. We
treat the primordial plasma as a fluid, and handle the dissipation of the baryon inhomogeneity through
hydrodynamic equations. This allows us to take into account the effects of mutual diffusion. We discuss
the hydrodynamics of the primordial plasma Section II. In Section III we present results from numerical
simulations. In the last two sections we compare the
predicted isotope yields with observations and give our conclusions.

Throughout this paper we use the natural unit system where $c=\hbar=k_B=1$.


\section{Dissipation of baryon inhomogeneity}

\subsection{Ions}

Consider the evolution of a density fluctuation in the baryonic component of the
primordial plasma. We are interested in the temperature range
$T\sim10$ MeV-1 keV.  We treat each isotope as a separate fluid. 
We write down the hydrodynamic equations for isotope $i$:
\begin{eqnarray}
     \frac{\partial n_i}{\partial t} &=& 
          -\nabla\cdot(n_i\ve v_i) 
          +\frac{\partial n_i}{\partial t} \vert_{\rm reac} \label{hydro1}\\
     \frac{\partial\ve q_i}{\partial t} &=&
         -T\nabla n_i +\sum_{j\ne i} \ve F_{ij} +\ve F_{ie} -n_iZ_i e\ve E 
            \label{hydro2}
\end{eqnarray} 
Here $n_i$ and $\ve q$ denote, respectively, the number and momentum density of isotope $i$. 
We have ignored second-order terms in fluid velocity $\ve v_i$, which is assumed
to be small. We use the non-relativistic formula for pressure
$P=nT$ and assume that temperature is nearly homogeneous. 
As pointed out in \cite{Jedamzik94a}, the fluctuations in temperature are of the order of the
baryon-to-photon ratio $\sim10^{-9}$. 

The last term in Eq. (\ref{hydro1}) represents production or destruction of ions via nuclear
rections.  Terms
$\ve F_{ie}$ and $\ve F_{ij}$ represent momentum transfer due to collisions on electrons or
other fluid components. The last term in Eq. (\ref{hydro2}) represents an electric field, which is
present if there is a departure from local charge neutrality.
In the following we evaluate explicit formulae for the collision terms
$\ve F$.
\bigskip

{\em Scattering between non-relativistic particles.}

The momentum transfer between two non-relativistic fluid components close to thermal
equilibrium is given by \cite{Present}
\begin{eqnarray}
       \ve F_{kj} =& n_k n_j \int\int \ve{d^3p}_k \ve{d^3p}_j \times\hbox{\hskip20mm} \nonumber \\
                   & f_k(\ve p_k)f_j(\ve p_j) \vert \ve u_j-\ve u_k\vert
                   \sigt_{kj}(p_{jk}) \ve p_{jk},   \label{nonrelf1}
\end{eqnarray}
where $f_k(\ve p)$ is the momentum distribution of particle $k$, 
such that $n_kf_k(\ve p)$ gives the phase space density,
$\ve u_j-\ve u_k$ is the relative velocity,
$\ve{p}_{jk}$ is the center-of-mass momentum,
and
\beq
          \sigt_{kj} = \int \frac{d\sigma_{kj}}{d\Omega}(1-\cos(\theta))d\Omega.
\eeq
is the transport cross section.
Assuming a small deviation from the Maxwellian distribution $f^0_k$,
\beq
      f_k(\ve p)=f_k^0(\ve p)\left( 1+\frac{\ve v_k\cdot\ve p}{T}\right),
\eeq 
where $\ve v_k=\langle u_k\rangle$ is the fluid velocity,
we obtain
\beq
     \ve F_{kj} =  -n_kn_j S_{kj} (\ve v_k-\ve v_j).
                                                         \label{nonrelf2}
\eeq
where 
\beq
      S_{kj} = \frac83 \left(\frac{2T\mu}{\pi}\right)^{1/2}
                                    \tilde\sigma_{kj}(T).   \label{smatrix}
\eeq
Here $\mu$ is the reduced mass
and the thermally averaged cross section is given by
\beq
            \tilde\sigma_{kj}(T) =
                  \frac{1}{(2\mu T)^3} \int_0^\infty
                  \exp\left(-\frac{k^2}{2\mu T}\right) k^5
                  \sigt_{kj}(k) dk,                   \label{effcross}
\eeq
where $k$ is the center-of-mass momentum.

Let us apply the above results to neutron-proton and ion-ion scattering.
At low energies (below a few MeV) the neutron-proton interaction is dominated
by s-wave scattering. The cross section is given by \cite{Preston}
\beq
    \sigma_{np}\!(k) \!=\! 
      \frac{ \pi a_s^2}{(a_sk)^2\!+\!(1\!-\!\half r_sa_sk^2)^2}
       \!+\! \frac{3\pi a_t^2}{(a_tk)^2\!+\!(1\!-\!\half r_ta_tk^2)^2} 
             \label{npcross}
\eeq
with $a_s=-23.71$ fm, $a_t=5.432$ fm, $r_s=2.73$ fm, and $r_t=1.749$ fm.
At zero-energy limit the cross section approaches the value 20440 mbarn.
The thermally averaged cross section (\ref{effcross}) can be evaluated numerically.

The transport cross section for Coulomb scattering between nonrelativistic
ions is given by
\beq
             \sigt = 4\pi (Z_iZ_j\alpha)^2\frac{\mu^2}{k^4}\Lambda
\eeq
where $\Lambda$ is the Coulomb logarithm \cite{LLkinetics,Goldston}.
The thermal cross section (\ref{effcross}) becomes
$\tilde\sigma=\pi(Z_iZ_j\alpha)^2\Lambda/(2T^2)$.
We then have
\beq
       S_{\rm ij} = \frac{4\pi}{3} \left(\frac{2\mu_{ij}T}{\pi}\right)^{1/2}
                                  \frac{(Z_iZ_j\alpha)^2\Lambda}{T^2}.
\eeq
\bigskip

{\em Scattering on electrons.}

The collisional force excerted on a heavy particle $k$ moving with velocity $\ve v_k$ through a
thermal background of light particles $j$ is given by
\beq
     \ve F_{kj} = n_k \int \ve{d^3p} \rho_j(\ve p)
                 \frac {\vert\ve p\vert}E
            \sigt_{kj}(\ve p)\ve p = -\frac{1}{b_{kj}} n_k\ve v_k. 
                                                      \label{relf}
\eeq
This equation relates the force to the mobility $b_{kj}$ \cite{LLkinetics,LLfluid}.
Here $\rho_j(\ve p)$  is the phase space density of particle
$j$ in the frame of particle $k$. Assuming a thermal distribution $\rho^0_j$ for
particle $j$ in laboratory frame, we have 
$\rho_j(\ve p)=\rho_j^0(E+\ve p\cdot\ve v)$, and
\beq
    -b_{ne}^{-1} = \int \ve{d^3p} \frac{d\rho^0_j(E)}{dE}
                   \frac {p^3}{3E} \sigt_{kj}(\ve p).
                  \label{mobility_def}
\eeq

Neutrons interact with electrons through their magnetic moment.
The transport cross section is \cite{AHS87}
\beq
            \sigt_{ne}
            = 3\pi\frac{\alpha^2\kappa^2}{m^2_n}
            = 8.07\times10^{-4}\mbox{ mbarn}.   \label{necross}
\eeq
where $\kappa=-1.91$ is the anomalous magnetic moment of the neutron.
Using MB statistics for electrons we obtain for the mobility
\beq
       b^{-1}_{ne} = \frac83\left(\frac{2m_e T}{\pi}\right)^{1/2}
                     \frac{K_{2.5}(z)}{K_2(z)}\sigt_{ne}n_e.
\eeq
where $K$ are modified bessel functions.

The differential cross section for a relativistic
electron scattering
on an ion with charge $Z$ is given by the Mott formula \cite{Lang}
\beq
             \frac{d\sigma}{d\Omega} =
             \frac{(Z\alpha)^2E^2_e}{4k^4\sin^4(\theta/2)}
             \left( 1-\beta^2\sin^2(\theta/2) \right).
                                                 \label{Mott}
\eeq
where the factor $1-\beta^2\sin^2(\theta/2)$ is the relativistic spin
correction. This gives the transport cross-section
\beq
          \sigt_{ie}(k) = 4\pi (Z\alpha)^2\frac{m^2_e+k^2}{k^4}\Lambda.
\eeq
where $\Lambda$ is the Coulomb logarithm.
Using again MB statistics we obtain
\beq
       b_{ie}^{-1} = \frac{4\pi T}{3}
                          \frac{(z^2+2z+2)}{K_2(z)e^z}
                          \frac{(Z_i\alpha)^2\Lambda n_e}{m^2_e}.
                                          \label{iedconst}
\eeq

\subsection{Electrons}

Thermal electron-positron pairs annihilate at temperatures $T\approx$1 MeV-20 keV. 
The remaining electrons must be treated as one fluid
component. 

For non-relativistic electrons we have
\begin{eqnarray}
     \frac{\partial n_e}{\partial t} &=& 
          -\nabla\cdot(n_e\ve v_e) \label{elhydro1}\\
     \frac{\partial\ve q_e}{\partial t} &=&
         -T\nabla n_e +\ve F_{e\gamma} +n_e e\ve E .
            \label{elhydro2}
\end{eqnarray} 
The term $\ve F_{e\gamma}$ represents Thomson scattering on background
photons \cite{Peebles},
\beq
   \ve F_{e\gamma} = -b_{e\gamma}^{-1}n_e\ve v_e
                   = -\frac43\sigma_T\epsilon_\gamma n_e\ve v_e, 
                    \label{Thomsondrag}
\eeq
where $\sigma_T=665$ mbarn is the Thomson cross section and $\epsilon_\gamma$ is the
energy density of background photons.

Note that formula (\ref{Thomsondrag}) is exactly valid only well after electron-positron
annihilation, when photon mean free path is large compared with the inhomogeneity scale. 
Around $T\approx20$ keV photons are still connected to the plasma. For precise treatment of this
transition period, photons should be included as one fluid component.

Ions diffusing out from the high-density regions leave behing a negative net charge.
That gives rise to an electric field, which forces electrons to move so as to restore the electrical
neutrality \cite{Spitzer}. Electrons are thus dragged along with
ions. The motion of ions is restricted by the Thomson drag force (\ref{Thomsondrag}), which acts on
them indirectly through the electric field.

If we assume spherical symmetry, the electric field at a given location is
determined by the total charge contained in the spherical region closer to the symmetry
center. The rate of change of the field is then
determined by the flux of charge through the sphere,
\beq
     \frac{\partial (e\ve E)}{\partial t} = 
       -4\pi\alpha\left( n_e\ve v_e-\sum_i n_iZ_i\ve v_i \right).
          \label{elfield}
\eeq

The five differential equations (\ref{hydro1}), (\ref{hydro2}), (\ref{elhydro1}), (\ref{elhydro2}), and
(\ref{elfield}), together with the formulae for momemtum transfer terms, form the basis of our
hydrodynamic simulations.

\subsection{Diffusion approximation}

It is interesting to look at how our hydrodynamic treatment relates to the common
diffusion approximation, where the evolution of inhomogeneity is presented by a differential
equation of the form
\beq
   \frac{\partial n}{\partial t} = \nabla\cdot(D\nabla n). \label{diffu}
\eeq

Consider the steady-state solution of Eq. (\ref{hydro2}) in absence of an electric field,
\beq
     -T\nabla n_i -\sum_{j\ne i}n_in_jS_{ij}(\ve v_i-\ve v_j) 
                  -b_{ie}^{-1}n_i\ve v_i = 0  \label{steady1}   
\eeq
If we ignore the motion of particle species other than $i$ ($v_j\approx0$ for $j\ne i$), 
equation (\ref{steady1}) and the continuity equation
(\ref{hydro1}) together lead to a diffusion equation of the form (\ref{diffu}), with
diffusion constants given by
$D_{ie}=b_{ie}T$ for scattering on electrons, and $D_{ij}=T/(n_jS_{ij})$ for
scattering between nuclei.

We note here that the neutron-proton and neutron-electron diffusion constants calculated this way
coincide with those given in \cite{Jedamzik01}. Also the proton-electron constant is in agreement at
the limit $\Lambda\gg1$.

The diffusion equation describes well the motion of a fluid if the
background fluid is stationary, so that its mutual motion can be ignored. We refer to this as
the approximation of independent diffusion. This approximation is valid in the case of diffusion
of neutrons, which are much more mobile than the ions and electrons they scatter on.  The
diffusion equation also describes well the motion of ions at high temperatures ($T\gg
20$ keV), where the dominant scattering process for ions is Coulomb scattering on background
electrons. Due to their large number the electrons can be regarded as  a stationary
background. At lower temperatures the situation is more complicated.  Ions gain or lose
momentum in collisions on other ion components, which move with comparable fluid velocities. 
Thus the mutual motion of the fluid components cannot be neglected.
The situation is further complicated by electron drag: electrons
are dragged along with ions so that charge neutrality is maintained.


\section{Simulations}

We have written an inhomogeneous nucleosynthesis code where a nuclear reaction
network is coupled to hydrodynamic equations. We assume spherical symmetry and use a non-uniform
radial grid of 64 cells. The grid is adjusted according to the density profile so that the cells are
smallest where the gradient of the baryon density is largest.  We assume a simple initial geometry with
a step-like density profile. The inner part of the simulation volume has a high baryon density $\eta_h$,
and the outer part a low density $\eta_l$. The initial conditions are determined by four
parameters: the volume fraction $f_v$ of the high-density region, density contrast
$R=\eta_h/\eta_l$, the average density $\eta=f_v\cdot\eta_h+(1-f_v)\cdot\eta_l$, and the
radius of the simulation volume $r$. We give $r$ in comoving units in meters at $T=1$ keV
temperature. One meter at $T=1$ keV corresponds to $4.26\times10^6$ m today.
The baryon density is given as the baryon-to-photon ratio $\eta$, which is related to $\Omega_b$
through
$\eta_{10}=10^{10}\eta=274\Omega_bh^2$.

The code evolves 21 variables in each grid zone: the concentration and
momentum density of $e$,
$n$, $p$, D,
${}^3$H, $\EHe$, $\UHe$, ${}^6$Li, $\ZLi$, and $\ZBe$, and the electric
field. These variables are evolved  
by solving a set of $21\times64$ stiff differential equations in time steps proportional to
the age of the universe. This involves the solution of a band diagonal linear system of rank
1344 at every time step. 
The nuclear reaction rates include those given in the NACRE compilation \cite{NACRE}.
The simulation is started at
$T=10$ MeV and ends at
$T=1$ keV, or when all nuclear reactions have ceased. 
The final output consist of the average
concentrations of
$p$, D, $\EHe$ (including $\EH$), $\UHe$,
${}^6$Li, and $\ZLi$ (including $\ZBe$).

For comparison we also made a set of simulations where we mimicked the approximation
of independent diffusion. We removed from the matrix all elements corresponding to mutual
diffusion, i.e, terms that represent dependence of $\partial\ve q_i/\partial t$ on $\ve
v_{j\ne i}$. We also forced a steady-state solution. The electron drag was taken into account
by adding to the momentum loss of an ion with charge $Z_i$ the Thomson force that would act on $Z_i$
electrons moving with the same velocity. This approach can be written as
\begin{eqnarray}
     && \frac{\partial n_i}{\partial t} =
          -\nabla\cdot(n_i\ve v_i) 
          +\frac{\partial n_i}{\partial t} \vert_{\rm reac} \label{hydro1cmp}\\
     && -T\nabla n_i -\sum_{j\ne i} n_in_jS_{ij}\ve v_i 
           -b^{-1}_{ie}n_i\ve v_i -b_{e\gamma}^{-1}Z_in_i\ve v_i = 0.
            \label{hydro2cmp}
\end{eqnarray}

\subsection{Separation of elements}

\begin{figure}[tbp]
\epsfxsize=8.0cm
\epsffile{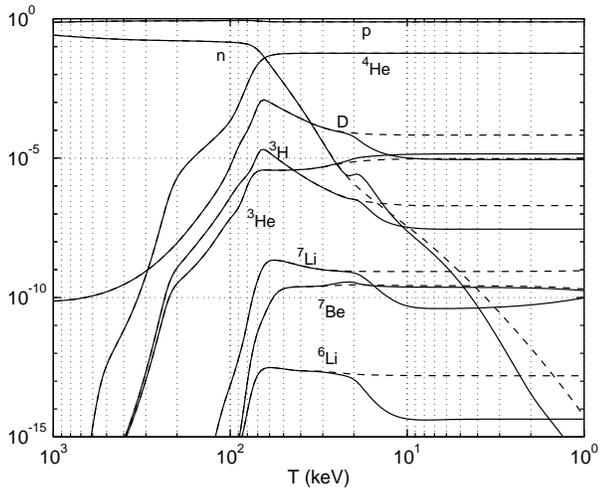}
\vskip 3mm
\caption[a]{
Evolution of the abundances of light isotopes as a function of temperature, for simulation 
parameters $r=10^7$ m (at 1 keV), $\eta_{10}=6$,
$f_v=0.01$, and $R=10^6$. The solid lines show results from a complete simulation. 
The dashed lines represent a simulation where the approximation of independent diffusion
was applied.  Separation of elements leads to a second burst of
nucleosynthesis below $T<20$ keV.}
\label{fig:isotopes}
\end{figure}

Figure
\ref{fig:isotopes} shows the light element abundances plotted against temperature, for
simulation parameters $r=10^7$ m, $\eta_{10}=6$, and $f_v=0.01$. 
We compare results from a complete simulation, and from a simulation where we used the approximation
(\ref{hydro2cmp}). The complete simulation shows a second burst of
nucleosynthesis at $T\approx20-10$ keV, leading to destruction of $\ZLi$, D, and $\EH$. 

This can be understood as follows.
Diffusion of electrons is inefficient at temperatures $T\gg1$ keV due to the frequent
Thomson scattering on background photons. Thomson drag then resists the diffusion of ions which must
drag electrons with them to maintain electrical neutrality. However, if we
divide the motion of ions into two components, one that obeys the condition $\sum_in_i\ve
v_iZ_i=0$ and thus carries no charge, and one that does carry charge, only the latter is resisted by
Thomson drag. Protons and helium ions, for instance, are allowed to diffuse into
opposite directions in such a way that the total charge flux vanishes.
This leads to a separation of elements. Heavier elements tend to get enriched in the
high-density regions, while protons diffuse out. The nucleosynthesis process in the
high-density region is enhanced by the increased concentration of heavier nuclei. 
This effect is responsible for the modified nucleosynthesis yields that our simulations show.

\begin{figure}[tbp]
\epsfxsize=8.0cm
\epsffile{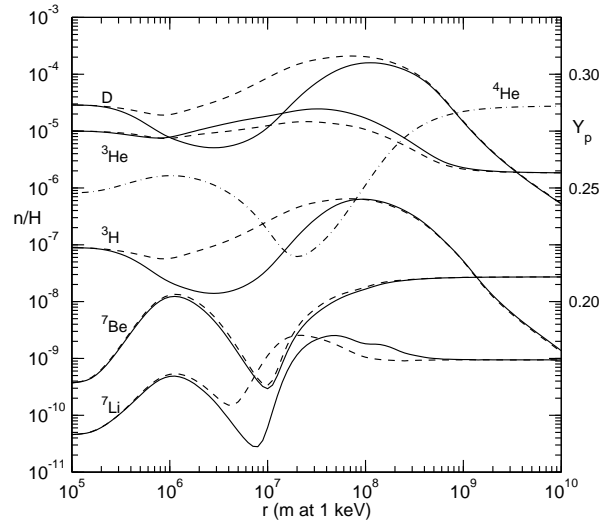}
\vskip 3mm
\caption[a]{
Light element abundances as a function of $r$, for simulation parameters $\eta_{10}=6$,
$f_v=0.01$, and $R=10^6$. The solid lines show results from a complete simulation. 
The dashed lines represent a simulation where mutual diffusion was ignored. 
The helium mass fraction $Y_p$ (dash-dotted line) is shown on linear scale (right y-axis). Other
isotopes are shown on logarithmic scale (left y-axis) as ratio of number density to that of hydrogen. }
\label{fig:risotopes}
\end{figure}

In Fig. {\ref{fig:risotopes}} we show the isotope yields as a function of inhomogeneity 
length scale, for $f_v=0.01$ and $\eta_{10}=6$. Again, we show results both
for a complete simulation (solid lines) and for a simulation with same parameters 
but with the approximation of independent diffusion.
The complete simulation shows a clear decrease in the abundances of D, $\EH$, and $\ZLi$, 
as compared to the diffusion approximation. 
Also the yields of $\EHe$ and $\ZBe$ are reduced, but not as
clearly. The yield of $\UHe$ is hardly affected.

Lithium is destroyed via reaction $\ZLi(p,a)\UHe$.
The mean destruction channel for $\ZBe$, instead, is via reaction $\ZBe(n,p)\ZLi$.
As this reaction requires free neutrons, which are not available after the
main phase of nucleosynthesis, $\ZBe$ is affected little in the second
nucleosynthesis phase. The same holds for $\EHe$, whose main reaction channel is
$\EHe(n,p)\EH$. 

The reduction in D, $\EH$ and $\ZLi$ due to element separation is most efficient at scales
somewhat smaller than the ``optimal scale'' at which the $\UHe$
yield is minimized. The maximal $\UHe$ reduction occurs at a scale at which neutrons diffuse
maximally out from the high-density regions, but the later back-diffusion is not too
efficient. At a somewhat smaller scale, back-diffusion of neutrons leads to synthesis
of nuclei in a narrow zone surrounding the high-density region. There are then plenty of
nuclei to be transported deeper into the high-density region, once the separation of
elements begins at $T\sim20$ keV.

Some analytic considerations may be in place here.
Consider the steady-state
solution of Eqs. (\ref{hydro2}) and (\ref{elhydro2}),
\begin{eqnarray}
     &-T\nabla n_i &-\sum_jn_in_jS_{ij}(\ve v_i-\ve v_j) 
      -n_iZ_i e\ve E = 0  \label{steady2}     \\
     &-T\nabla n_e &-b_{e\gamma}^{-1} n_e\ve v_e +n_ee\ve E = 0 . \label{steady3}
\end{eqnarray} 
We are interested in the temperature regime $T<20$ keV and have ignored terms that represent
scattering on electrons. A measure of the distance over which $n_e$ can deviate from
$\sum_in_iZ_i$ is given by the Debye shielding distance $h=(4\pi\alpha n_e/T)^{-1/2}$ \cite{Spitzer},
which is orders of magnitude smaller than the inhomogenity length scale. We can therefore assume that
electrical neutrality holds at the scale of the inhomogeneity ($n_e\approx\sum_in_iZ_i$), and based on
that eliminate the field $\ve E$.

The evolution of the inhomogeneity is particularly simple in two
limiting cases. If the interaction between ions is strong compared with the electron-photon
interaction ($n_iS_{ij}\gg b_{e\gamma}^{-1}$), as is the case at late times ($T\ll 1$ keV), the plasma
behaves as a single fluid, moving with a collective velocity
$\ve v_e$. Taking the sum of Eqs. (\ref{steady2}) and (\ref{steady3}) and using the electrical
neutrality we find
\beq
      -T(\nabla n_B+\nabla n_e) - b_{e\gamma}^{-1} n_e\ve v_e = 0.
\eeq
This represents collective hydrodynamic motion of the plasma against Thomson drag
\cite{Alcock90,Jedamzik94a}. With the approximation $n_b\approx n_e$ it leads to the
diffusion equation for the baryon density, with diffusion constant $D_{hyd}=2b_{e\gamma}T$. 

In the opposite limiting case, when electron-photon scattering dominates over ion-ion
scattering, the motion of electrons is suppressed by the Thomson drag. Ions then move under
the condition that the net current carried by ions vanishes, $\sum_in_iZ_i\ve v_i=0$.
Consider for simplicity a system of two ion species only, say hydrogen and $\UHe$.
The steady-state solution now
simplifies into
\beq
    T\left( Z_1\frac{\nabla n_2}{n_2}-Z_2\frac{\nabla n_1}{n_1}\right)
     = (\ve v_1-\ve v_2)S_{12}(n_1Z_1+n_2Z_2).
\eeq
It is now easy to see that if two isotopes have the same initial inhomogeneity ($\nabla
n_1/n_1=\nabla n_2/n_2$), then the one with smaller charge will flow into the direction of
negative density gradient, while the one with larger charge will move into the opposite
direction. The isotope with larger charge will get concentrated into the
high-density region.

\begin{figure}[tbp]
\centerline{\epsfxsize=8.5cm
\epsffile{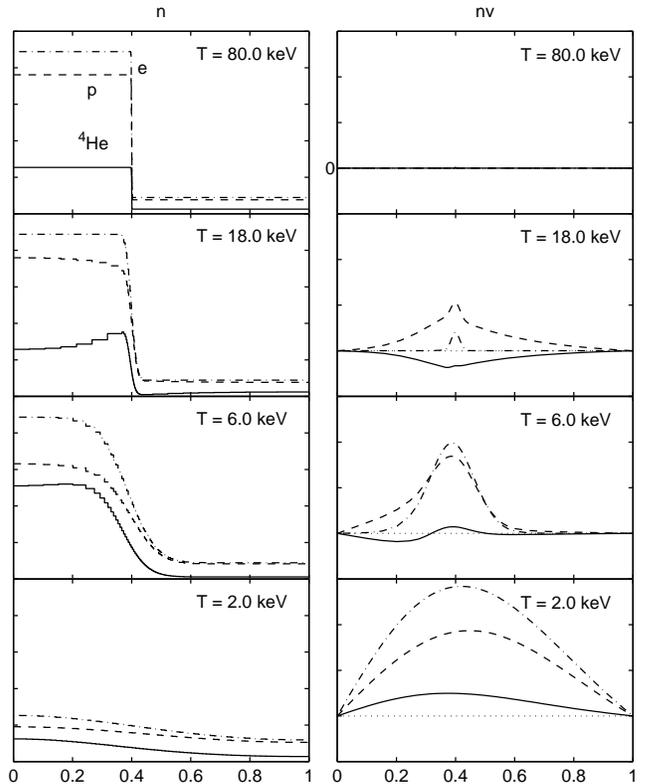}}
\vskip 3mm
\caption[a]{
Separation of elements. Number density (left) and $n\ve v$ (right) are shown for a
  run with with only protons and $\UHe$ present. Nuclear reaction were turned off.
  The simulation started with an initial profile with $R=10$, $r=10^7$ m, $f_v=0.4^3$, and
 uniform helium mass fraction
  $Y_p=0.25$ (uppermost frame). We show the density profile and flux of $\UHe$ (solid line), $p$
(dashed line), and
$e$ (dash-dotted line) along the radial axis of the sperical simulation volume. Helium
begins to concentrate in the high-density region at
$T<20$ keV. The concentration of $\UHe$ in the center reaches a maximum around $T\approx 6$ keV. The
inhomogeneity is finally erased by collective hydrodynamic expansion against Thomson drag (last frame).
}
\label{fig:art}
\end{figure}

In order to illustrate the separation of elements, we made a run 
with only protons and $\UHe$ present. We started with a
step-like initial profile with uniform helium mass fraction
$Y_p=0.25$. The concentration profiles of the two elements and electrons, as well as density times
velocity, at various temperatures are shown in Fig. \ref{fig:art}. Helium begins to concentrate in the
high-density region at $T<20$ keV. At $T=6$ keV the initial
high-density region contains 62\% of all helium nuclei, while in the beginning it contained 41\%.
The inhomogeneity is erased when Thomson scattering becomes inefficient in restricting the
collective motion of the plasma.


\section{IBBN computations and confrontation with observations}

The best way to evaluate the primordial $\ZLi$ has for a long time been the so-called Spite plateau
\cite{Spite} observed in halo stars. 
There is still debate on how much the $\ZLi$ in
Spite plateau stars has depleted from the primordial abundance, and consequently, on the primordial
$\ZLi$ abundance. While some authors obtain an relatively high upper limit
$\ZLiH<4\times10^{-10}$\cite{Pins}, a number authors
argue for a lower value\cite{Ryan99,Ryan00,Suzuki00}. 
Ryan \etal\cite{Ryan00} derive the range
$-10.04<\log_{10}(\ZLiH)<-9.72$ for the primordial abundance. In SBBN this corresponds to
$\eta_{10}=10^{10}\eta<4.2$. Suzuki \etal obtain an even tighter range $-9.97<\log_{10}(\ZLiH)<-9.77$.
A recent study \cite{Bonifacio02} gives an intermediate estimate
$\log_{10}(\ZLiH)=-9.76\pm0.056\pm0.06$.  

The tight lithium limits of Ryan \etal and and Suzuki \etal are in
conflict with the low deuterium estimations from QSO absorption systems
\cite{OMeara01,DOdorico01,Pettini01}. O'Meara \etal obtain the range $\DH=(3.0\pm0.4)\times10^{-5}$
from a combined study of four such systems. In SBBN this corresponds to $5.4<\eta_{10}<6.4$. 
The tight lithium limits are also in conflict with the recent Boomerang result
$\eta\sim6\times10^{-10}$ \cite{boomerang}.

In light of the above,
it is interesting to note that the separation of elements due to Thomson drag leads to
simultaneous destruction of
$\ZLi$ and D. 

\begin{figure*}[tp]
\vbox{
\centerline{\hbox{
\epsfxsize=80mm\epsffile{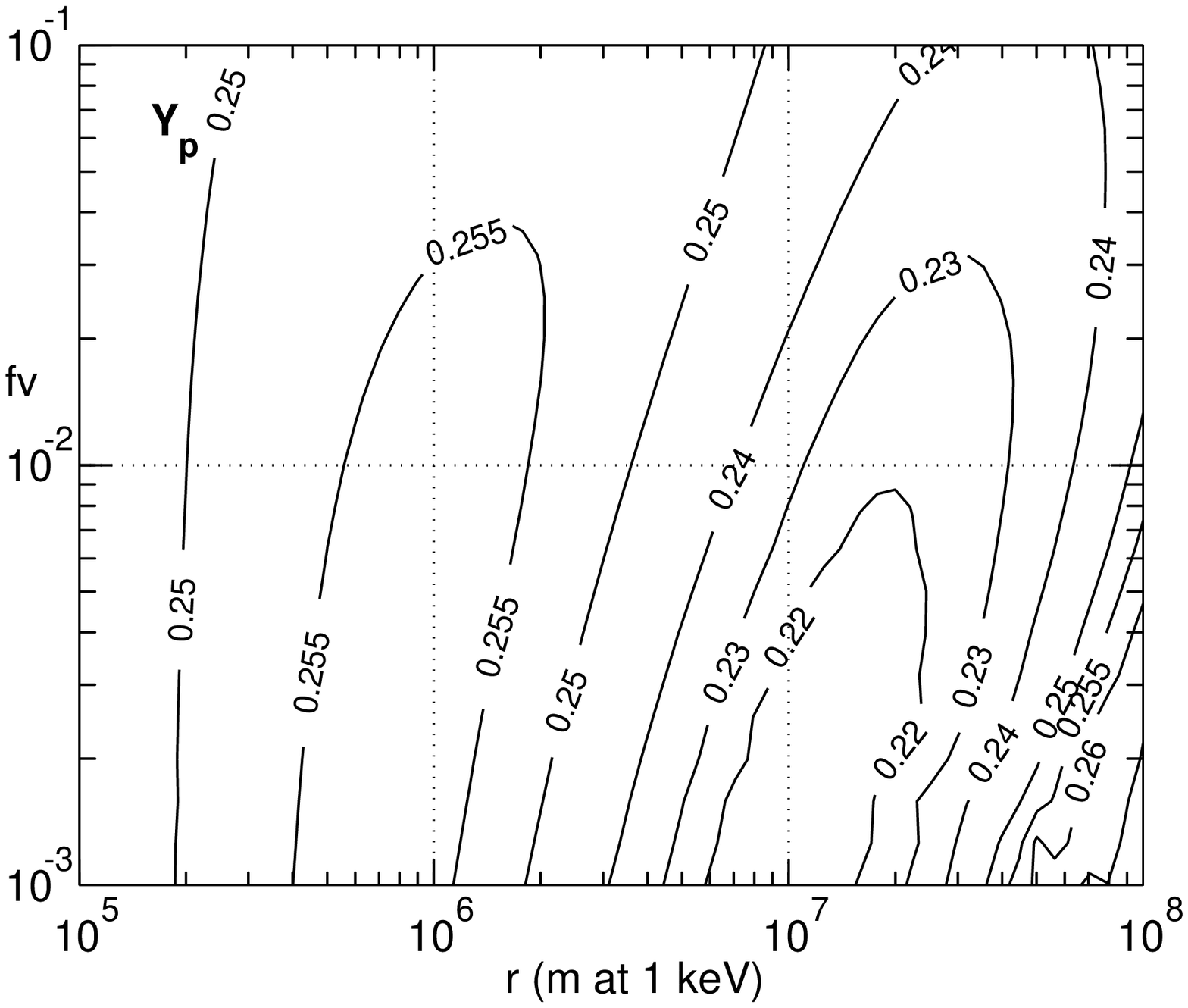}
\epsfxsize=80mm\epsffile{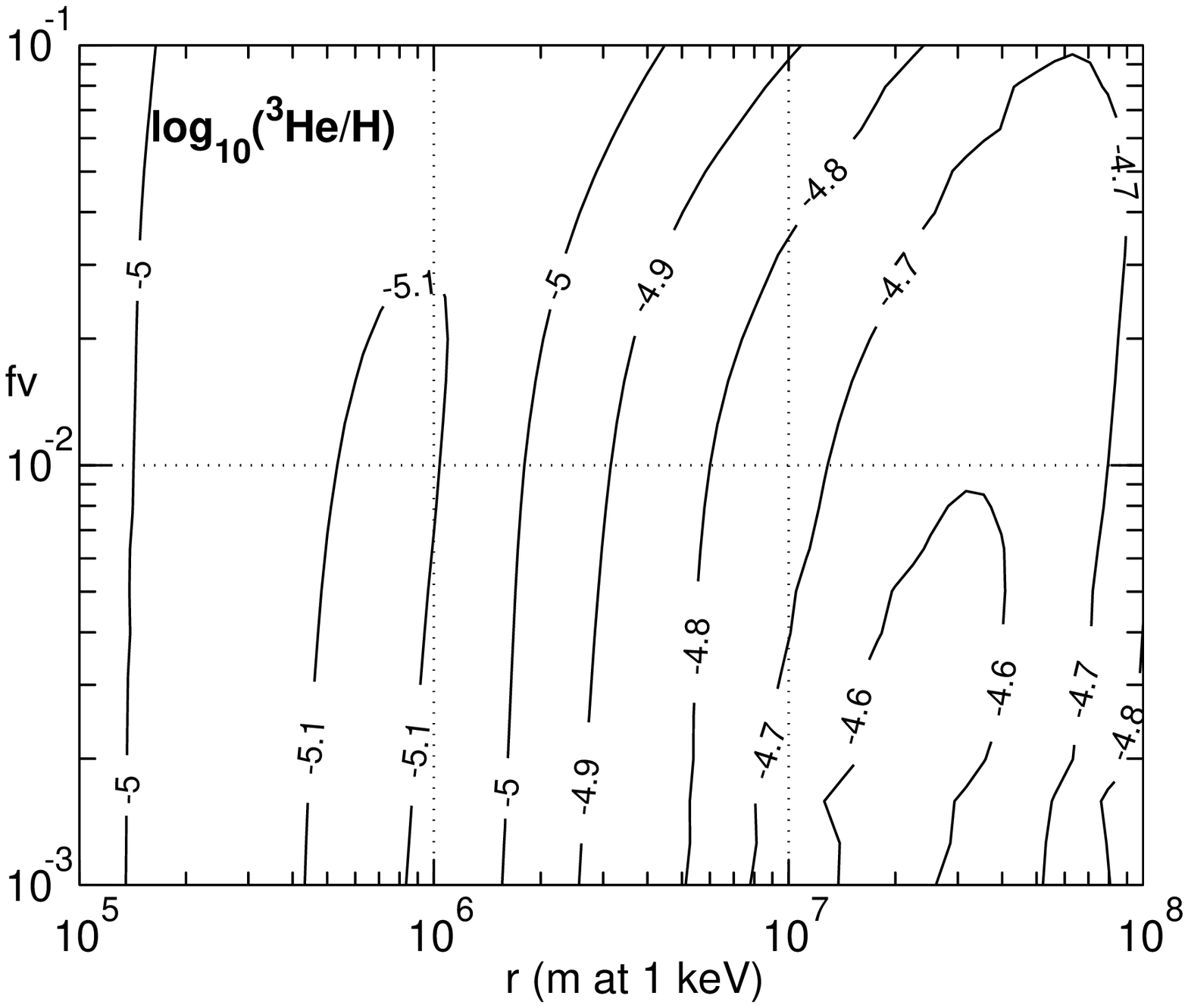}}}
\centerline{\hbox{
\epsfxsize=80mm\epsffile{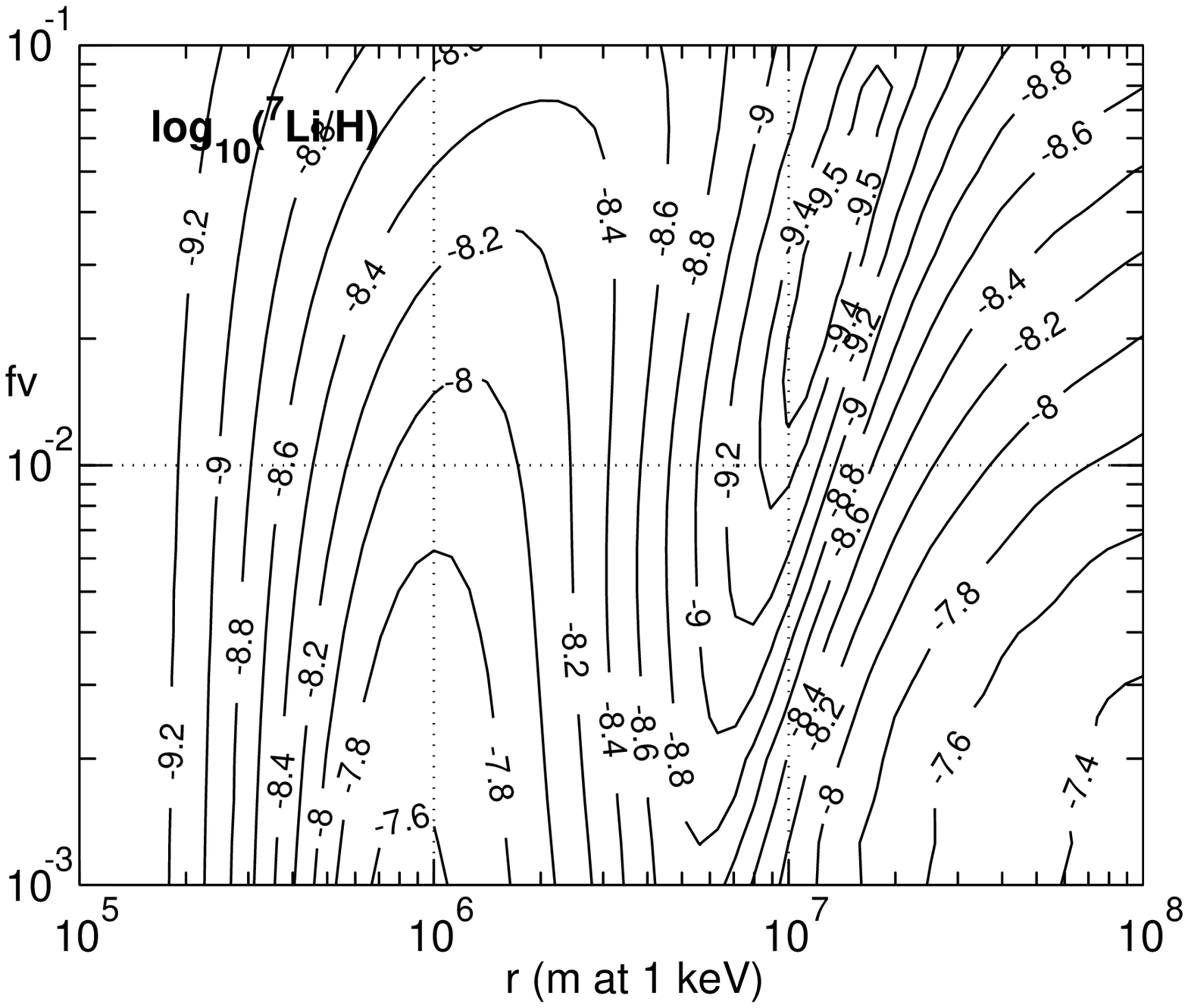}
\epsfxsize=80mm\epsffile{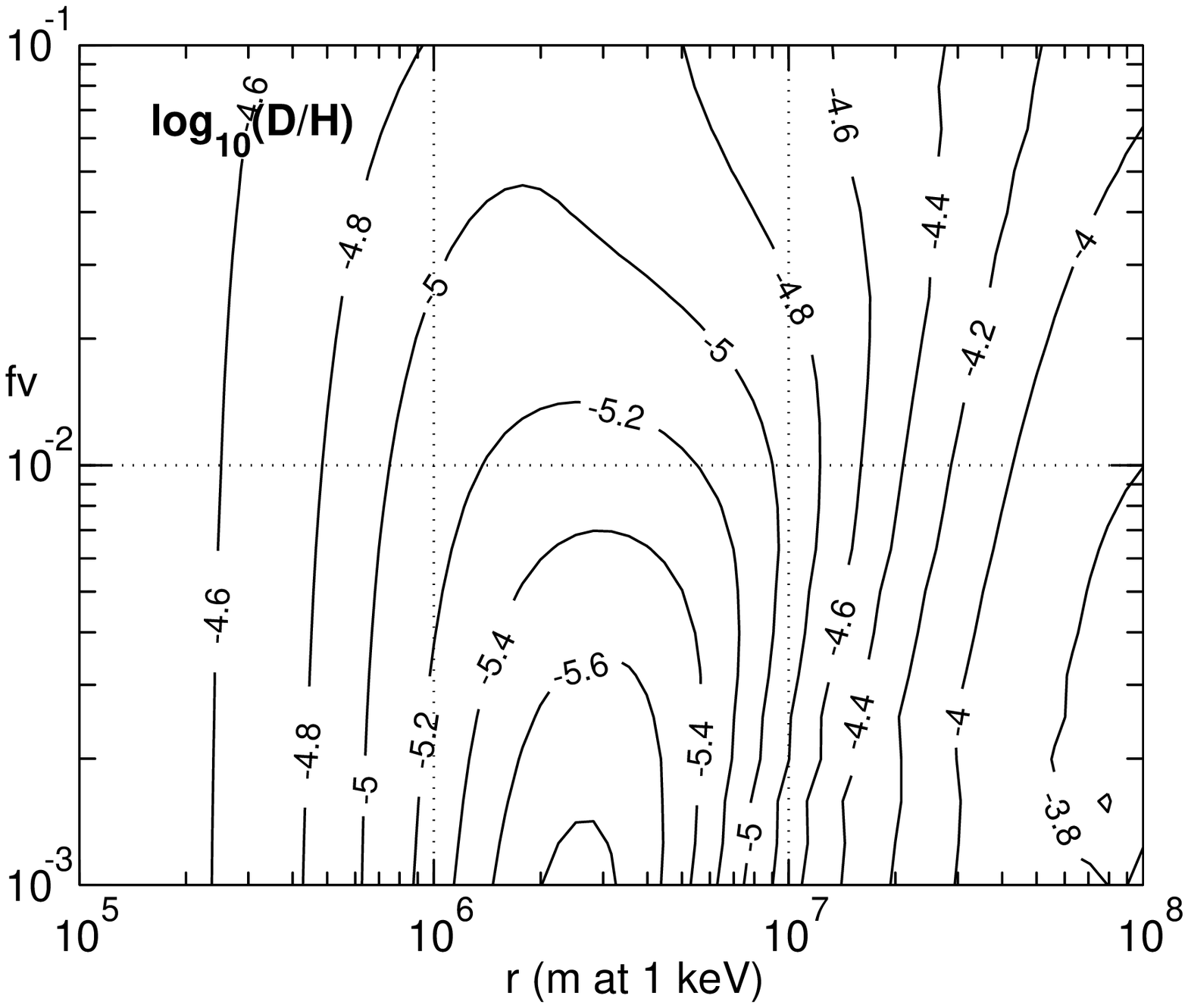}}}
\vskip 5mm
}
\caption[a]{
Isotope yields as a function of the inhomogeneity length scale $r$ and the volume fraction $f_v$ of the
high-density region. Other simulation parameters were $\eta=6\times10^{-10}$ and $R=10^6$. 
At the left the
results converge towards SBBN yields $\DH=2.9\times10^{-5}$,
$\EHeH=1.0\times10^{-5}$, $Y_p=0.2483$, and $\ZLiH=4.1\times10^{-10}$.}
\label{fig:rfv}
\end{figure*}

We have computed the light element abundances for a range of
inhomogeneity parameters.
Figure \ref{fig:rfv} shows the yields of light isotopes as a function of length scale $r$ and 
volume fraction of the high-density region $f_v$, for $\eta=6\times10^{-10}$. At small scales the
results converge towards SBBN results.
The smallest $\ZLi$ yield $\ZLiH=10^{-9.55}$ was obtained at $f_v\approx10^{-1.5}\approx0.032$ and
$r\approx7.1$ m. The SBBN value is $\ZLiH=10^{-9.39}$.

\begin{figure*}[tp]
\vbox{
\centerline{\hbox{
\epsfxsize=80mm\epsffile{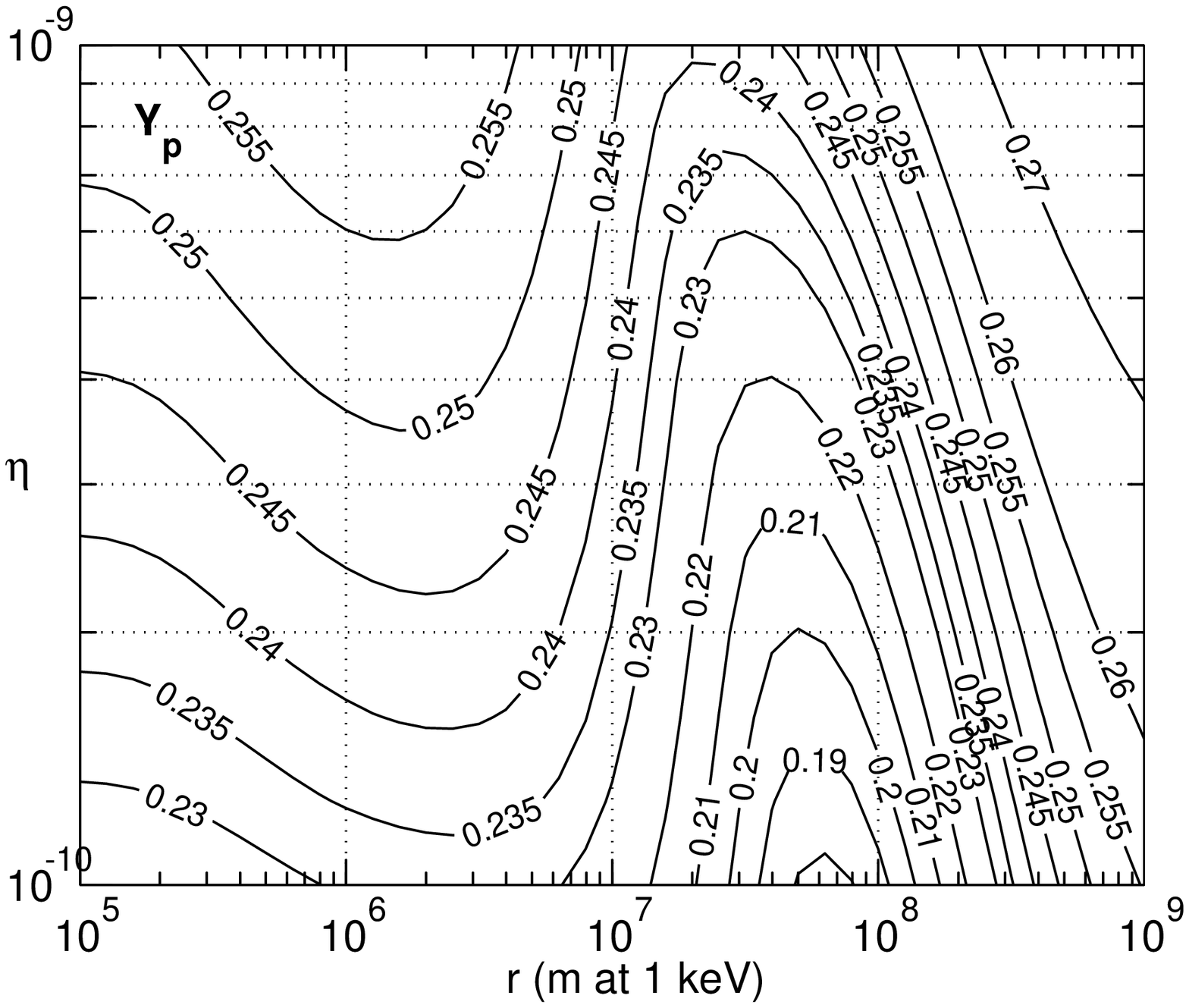}
\epsfxsize=80mm\epsffile{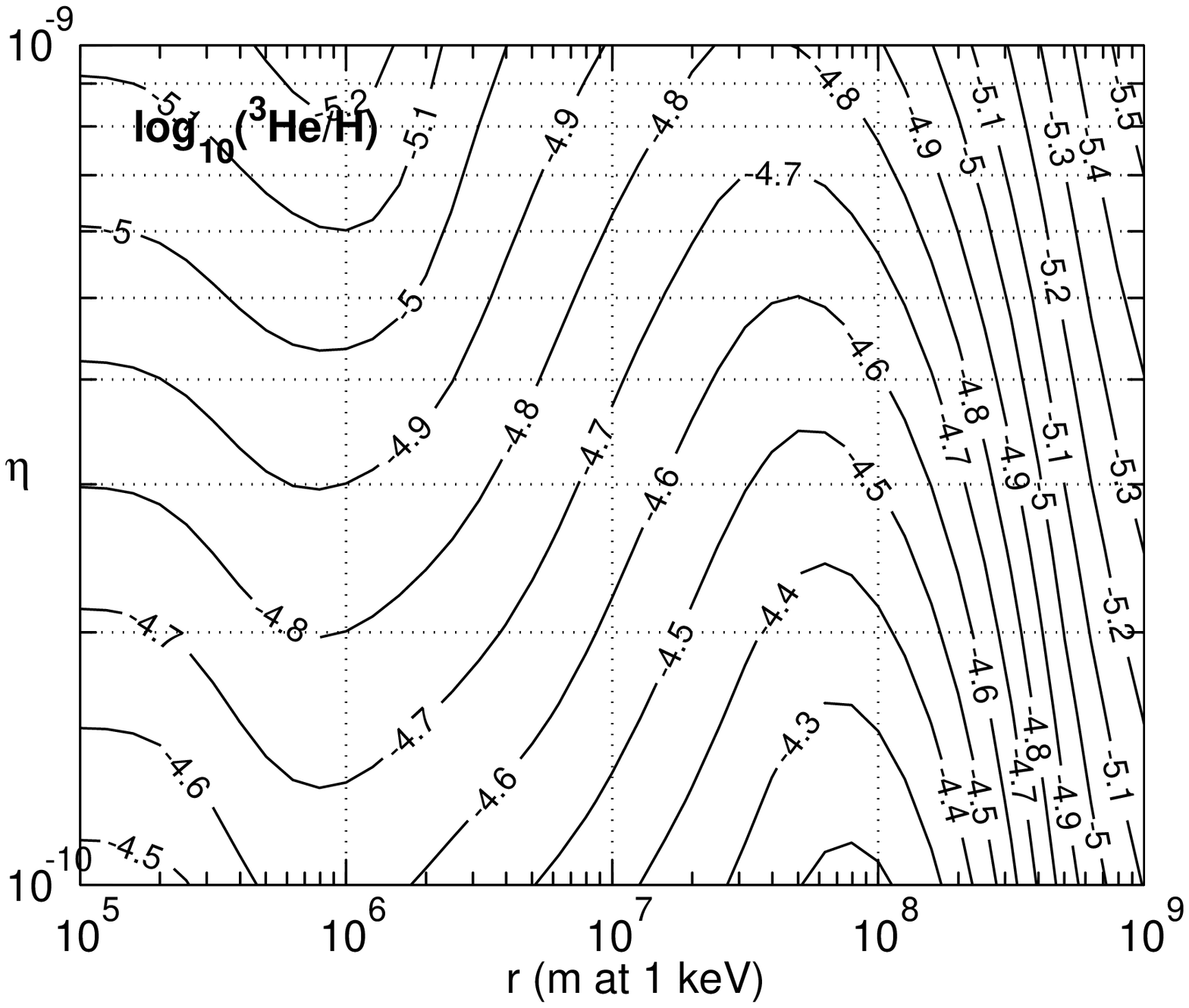}}}
\centerline{\hbox{
\epsfxsize=80mm\epsffile{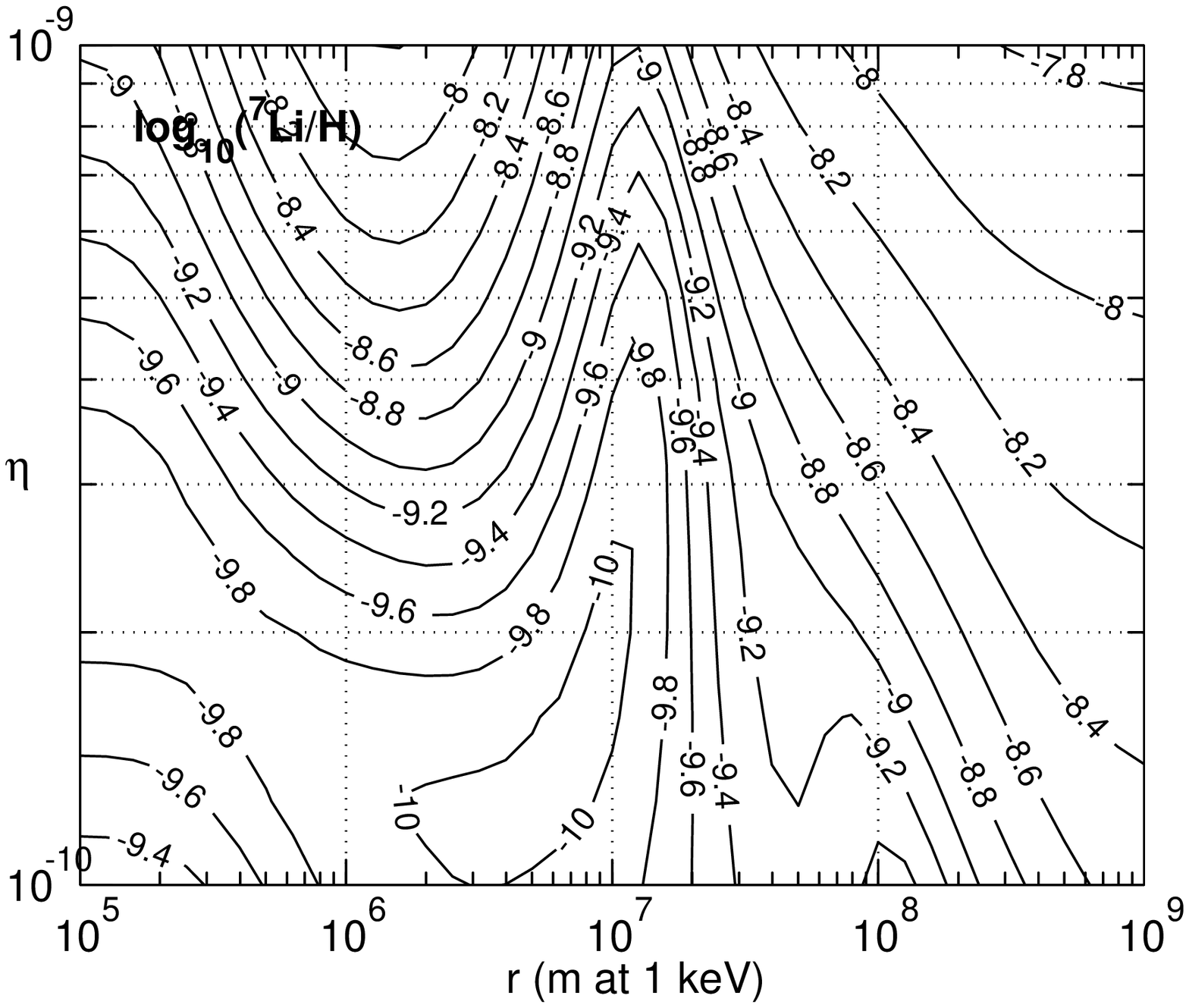}
\epsfxsize=80mm\epsffile{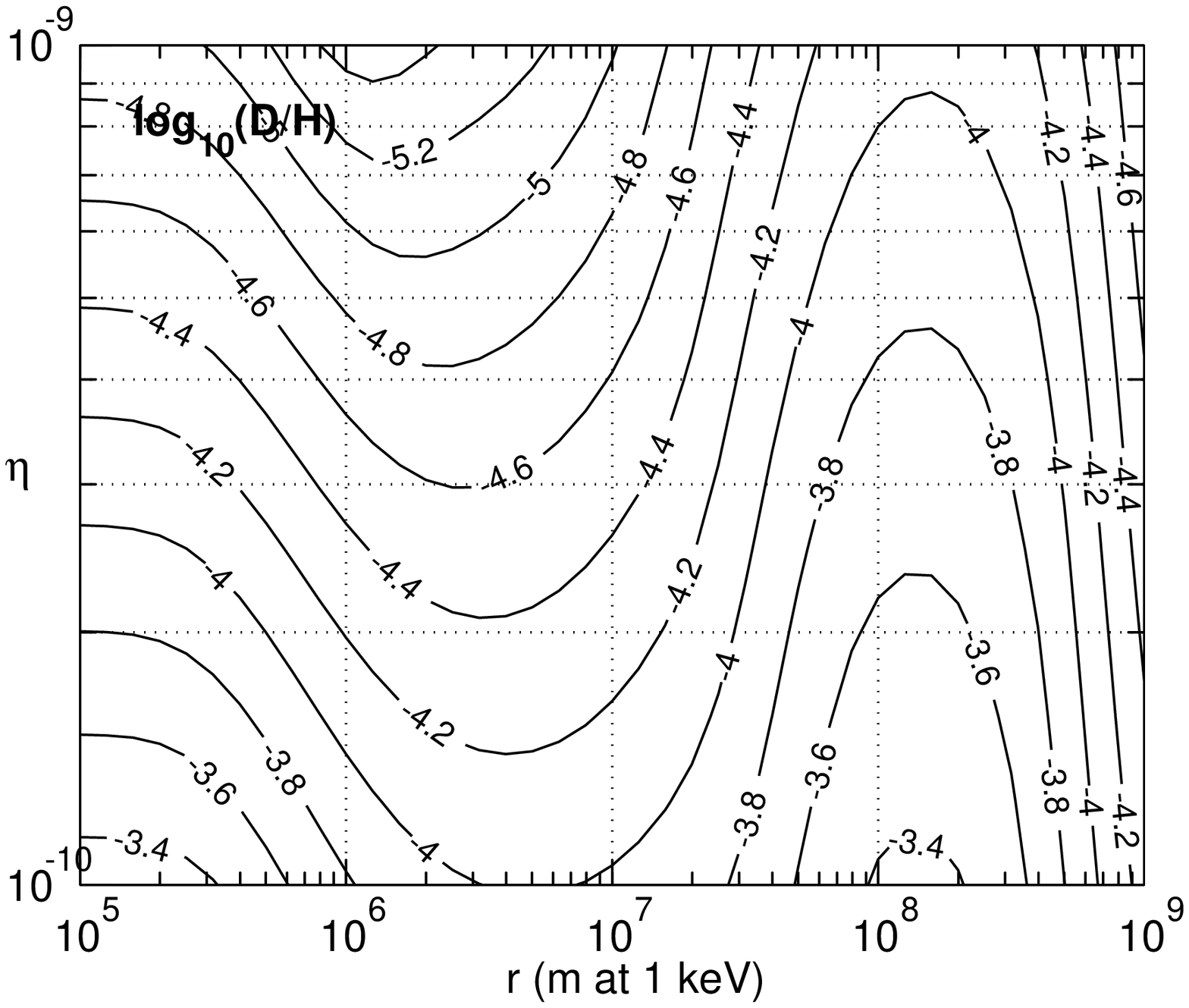}}}
\vskip 5mm
}
\caption[a]{
Isotope yields on $(r,\eta)$ plane for $f_v=10^{-1.5}$ and $R=10^6$.
}
\label{fig:reta}
\end{figure*}

In Fig. (\ref{fig:reta}) we show the isotope yields as a function of $r$ and $\eta$, for $f_v=10^{-1.5}$
and $R=10^6$.
The reduction in $\ZLiH$ is more prominent at low $\eta$, due to the fact that at low $\eta$ most of
the lithium is produced directly as $\ZLi$, which is sensitive to the separation of elements. At high
$\eta$ most of the lithium comes from $\ZBe$, which is not affected as much.

\begin{figure}[tbp]
\vbox{
\epsfxsize=80mm
\epsffile{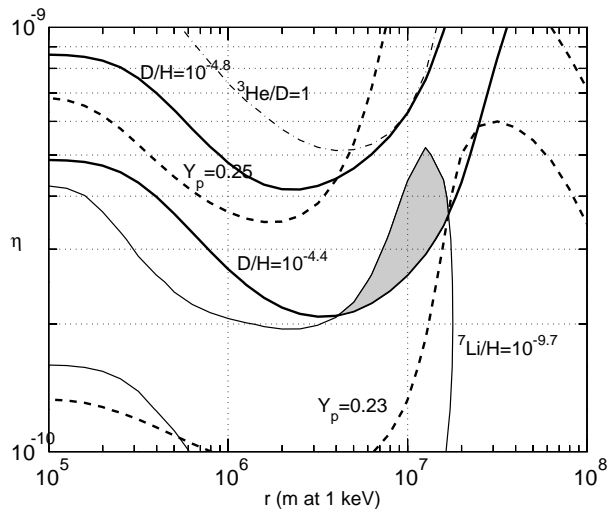}
\vskip 1mm
}
\caption[]{
Observational constraints on $(r,\eta)$ plane.
The shaded region satisfies the constraints $0.23<Y_p<0.25$ (thick dashed lines),
$10^{-4.8}<\DH<10^{-4.4}$ (thick solid lines), $\ZLiH<10^{-9.7}$ (thin solid line), and $\EHe/{\rm
D}<1$ (dash-dotted line). }
\label{fig:combined}
\end{figure}

In Fig. (\ref{fig:combined}) we compare our results to a set of observational constraints on light
element abundances. The simulation parameters are the same as in Figs. (\ref{fig:reta}). 
For $\UHe$ we choose conservative limits $0.23<Y_p<0.25$. For deuterium we select constraints
$10^{-4.8}<\DH<10^{-4.4}$. The lower limit here comes from the present $\DH$ abundance in interstellar
medium \cite{Linsky}. The upper limit is based on the two-sigma upper limit of the O'Meara estimation
\cite{OMeara01}.
For lithium we choose a low limit
$\ZLiH<10^{-9.7}$. We also include the constraint $\EHe/{\rm D}<1$. The lithium and deuterium
constraints we have chosen are in conflict in SBBN.  The upper limit to deuterium implies
$\eta_{10}>4.8$ while the upper limit to $\ZLi$ implies $\eta_{10}<4.2$. In IBBN we find a region in
the parameter space, where all constraints are satisfied simultaneously. The allowed region falls in
the range
$2.1<\eta_{10}<5.2$, corresponding to $0.008<\Omega_bh^2<0.019$. We note that if we apply the
approximation of independent diffusion, the allowed region disappears.

\section{Conclusions}

We have studied inhomogeneous big bang nucleosynthesis with emphasis on transport phenomena. We
combined a hydrodynamic treatment to a nucleosynthesis simulation. We found an effect that to
our knowledge has been overlooked before: separation of elements due to Thomson drag. Thomson drag
prevents the diffusion of the electron fluid shortly after
electron-positron annihilation. Hydrogen and multiply charged elements then diffuse into opposite
directions so that no net charge is carried. Helium and lithium get concentrated into high-density
regions, which leads to enhanched nucleosynthesis and destruction of $\ZLi$, D,
and $\EHe$. The effect is important at length scales from $10^5$ to $10^9$ meters at 1 keV
temperature, corresponding to $10^{-5}-0.1$ pc today.

We computed the light element yields for a variety of initial inhomogeneity profiles and found a
region in the parameter space where a low lithium constraint $\ZLiH<10^{-9.7}$ and a low deuterium
constraint $\DH<10^{-4.4}$ are satisfied simultaneously for $\eta=(2.1-5.2)\times10^{-10}$.


\section*{Acknowledgements}

I thank the Center for Scientific Computing
(Finland) for computational resources.



\begin{thebibliography}{99}

%
%

\bibitem[*]{maile}
Electronic address: Elina.Keihanen@helsinki.fi

%
%


\bibitem{Applegate85} J.H.~Applegate and C.J.~Hogan,
\PRD{31}{3037}{1985}.

\bibitem{AHS87} J.H. Applegate, C.J. Hogan, and R.J. Scherrer,
\PRD{35}{1151}{1987}.

\bibitem{Alcock87}  C.~Alcock, G.M.~Fuller, and G.J.~Mathews,
\ApJ{320}{439}{1987}.

\bibitem{Malaney88}  R.M.~Malaney and W.A.~Fowler,
\ApJ{333}{14}{1988}.

\bibitem{Terasawa89}  N.~Terasawa and K.~Sato,
\PRD{39}{2893}{1989}.

\bibitem{HX88}  H.~Kurki-Suonio, R.A.~Matzner, J.M.~Centrella,
T.~Rothman, and J.R.~Wilson,
\PRD{38}{1091}{1988}.

\bibitem{HX89}  H.~Kurki-Suonio and R.A.~Matzner,
\PRD{39}{1046}{1989}.

\bibitem{HX90a}  H.~Kurki-Suonio, R.A.~Matzner, K.A.~Olive, and D.N.~Schramm,
\ApJ{353}{406}{1990}.

\bibitem{HX90b}  H.~Kurki-Suonio and R.A.~Matzner,
\PRD{42}{1047}{1990}.

\bibitem{Mathews90}  G.J.~Mathews, B.S.~Meyer, C.R.~Alcock, and G.M.~Fuller,
\ApJ{358}{36}{1990}.

\bibitem{Mathews93}  G.J.~Mathews, D.N.~Schramm, and B.S.~Meyer,
\ApJ{404}{476}{1993}.

\bibitem{Jedamzik94b}  K.~Jedamzik, G.M.~Fuller, and G.J.~Mathews,
\ApJ{423}{50}{1994}.

\bibitem{Mathews96}  G.J.~Mathews, T.~Kajino, and M.~Orito,
\ApJ{456}{98}{1996}.

\bibitem{Orito97}  M.~Orito, T.~Kajino, R.N.~Boyd, and G.J.~Mathews,
\ApJ{488}{515}{1997}.

\bibitem{HX97}  H.~Kurki-Suonio, K.~Jedamzik, and G.J.~Mathews,
\ApJ{479}{31}{1997}.

\bibitem{Kainulainen99} K. Kainulainen, H. Kurki-Suonio and E. Sihvola,
\PRD{59}{083505}{1999}.

\bibitem{HX01}  H. Kurki-Suonio and E. Sihvola,
\PRD{63}{083508}{2001}.

\bibitem{Jedamzik01}  K.~Jedamzik and J.B.~Rehm,
\PRD{64}{023510}{2001}.


\bibitem{Alcock90} C.R. Alcock, D.S. Dearborn, G.M. Fuller, G.J. Mathews, and B.S. Meyer,
\PRL{64}{2607}{1990}.

\bibitem{Jedamzik94a}  K. Jedamzik and G.M. Fuller, \ApJ{423}{33}{1994}.

\bibitem{Present} See, e.g., 
R.D.~Present, {\it Kinetic Theory of Gases}, 
(McGraw-Hill Book Company, New York, 1958).



\bibitem{Preston}  M.A. Preston and R.K. Bhaduri,
{\it Structure of the Nucleus} (Addison-Wesley, Reading, MA, 1975).

\bibitem{LLkinetics}  L.D. Landau and E.M. Lifschitz,
{\it Physical Kinetics} (Pergamon, New York, 1981).

\bibitem{Goldston}  R.J. Goldston and P.H. Rutherford,
{\it Introduction to Plasma Physics} (IOP, Bristol,1995).

\bibitem{LLfluid}  L.D. Landau and E.M. Lifschitz,
{\it Fluid Mechanics} (Pergamon, New York, 1959).

\bibitem{Lang} K.R. Lang, {\it Astrophysical formulae} (Springer, Berlin,1980).

\bibitem{Peebles}  P.J.E. Peebles, {\it Physical Cosmology}
(Princeton University, Princeton, NJ, 1971);
P.J.E. Peebles, {\it Principles of Physical Cosmology}
(Princeton University, Princeton, NJ, 1993).

\bibitem{Spitzer} L.~Spitzer, Jr., {\it Physics of Fully Ionized Gases}
(Interscience, New York, 1956).




\bibitem{NACRE} C. Angulo \etal
(The NACRE collaboration), \np{A656}{3}{1999}.




\bibitem{Spite} M. Spite and F. Spite, \anap{115}{357}{1982};
\Nat{297}{483}{1982}; P. Bonifacio and P. Molaro,
\MNRAS{285}{847}{1997}.

\bibitem{Pins} M.H. Pinsonneault, T.P. Walker, G. Steigman, and
V.K. Narayanan, \ApJ{527}{180}{1999};  
M.H. Pinsonneault, G. Steigman, T.P. Walker, and V.K. Narayanan, astro-ph/0105439 (2001).

\bibitem{Ryan99} 
S.G. Ryan, J.E. Norris, and T.C. Beers, \ApJ{523}{654}{1999}.

\bibitem{Ryan00} 
S.G. Ryan, T.C. Beers, K.A. Olive, B.D. Fields, and J.E. Norris, \apjl{530}{L57}{2000}.

\bibitem{Suzuki00} 
T.K. Suzuki, Y. Yoshii, and T.C. Beers, \ApJ{540}{99}{2000}.

\bibitem{Bonifacio02} P. Bonifacio \etal, astro-ph/0204332 (2002). 



\bibitem{OMeara01} 
J.M. O'Meara, D. Tytler, D. Kirkman, N. Suzuki, J.X. Prochaska, D. Lubin, and A.M. Wolfe,
\ApJ{552}{718}{2001}.

\bibitem{DOdorico01} S. D'Odorico, M. Dessauges-Zavadsky, and P. Molaro,
astro-ph/0102162.

\bibitem{Pettini01} M. Pettini and D.V. Bowen, \ApJ{560}{41}{2001}.

\bibitem{boomerang} 
P. de Bernardis \etal, \ApJ{564}{559}{2002}.

\bibitem{Linsky} J.L. Linsky \etal, \ApJ{402}{694}{1993};
J.L. Linsky \etal, \ApJ{451}{335}{1995}.


\end{thebibliography}
\end{document}